\documentclass[letterpaper]{article} %
\usepackage{aaai24}  %
\usepackage{times}  %
\usepackage{helvet}  %
\usepackage{courier}  %
\usepackage[hyphens]{url}  %
\usepackage{graphicx} %
\usepackage{natbib}  %
\usepackage{caption} %
\usepackage{algorithm}
\usepackage{algorithmic}

\usepackage{newfloat}
\usepackage{listings}
\DeclareCaptionStyle{ruled}{labelfont=normalfont,labelsep=colon,strut=off} %
\floatstyle{ruled}
\newfloat{listing}{tb}{lst}{}
\floatname{listing}{Listing}
\nocopyright %

\usepackage{todonotes}
\usepackage{enumitem} 
\usepackage{mathtools}
\usepackage{amsthm}
\usepackage{amsmath}
\usepackage{amssymb}
\usepackage{tcolorbox}
\usepackage{graphicx}
\usepackage{subcaption}
\usepackage{soul}

\usepackage{proof}
\usepackage{tikz}
\usetikzlibrary{shapes.geometric, arrows.meta, trees}

\usepackage{subcaption}

\newcommand{\citeN}[1]{\citeauthor{#1} \shortcite{#1}}
\newcommand{\A}{\mathcal{A}}
\newcommand{\M}{\mathcal{M}}
\newcommand{\N}{\mathcal{N}}
\newcommand{\UF}{\textbf{UF}}
\newcommand{\DT}{\textbf{ADT}}
\newcommand{\LIA}{\textbf{LIA}}

\mathchardef\mhyphen="2D

\newtheoremstyle{thrmstyle}
  {10pt} %
  {10pt} %
  {\em} %
  {} %
  {\bfseries} %
  {.} %
  {.5em} %
  {} %
\theoremstyle{thrmstyle}
\newtheorem{theorem}{Theorem}[section]

\newtheorem{defi}{Definition}[section]

\title{An Eager Satisfiability Modulo Theories Solver for Algebraic Datatypes
\thanks{We would like to thank Adwait Godbole, Ameesh Shah, Jiwon Park and Shangyin Tan for their insightful feedback. This work was supported in part by a Qualcomm Innovation Fellowship, NSF grant 1837132, DARPA contract FA8750-20-C-0156, an Amazon Research Award, Toyota under the iCyPhy center, a UC Berkeley Summer Undergraduate Research Fellowship, and by Intel under the Scalable Assurance program.}
}
\author{
    Amar Shah,
    Federico Mora,
    Sanjit A. Seshia
}
\affiliations{
    University of California, Berkeley
}

\begin{document}

\maketitle

\begin{abstract}
Algebraic data types (ADTs) are a construct classically found in functional programming languages that capture data structures like enumerated types, lists, and trees. In recent years, interest in ADTs has increased. For example, popular programming languages, like Python, have added support for ADTs. Automated reasoning about ADTs can be done using satisfiability modulo theories (SMT) solving, an extension of the Boolean satisfiability problem with constraints over first-order structures. Unfortunately, SMT solvers that support ADTs do not scale as state-of-the-art approaches all use variations of the same \emph{lazy} approach. In this paper, we present an SMT solver that takes a fundamentally different approach, an \emph{eager} approach. Specifically, our solver reduces ADT queries to a simpler logical theory, uninterpreted functions (UF), and then uses an existing solver on the reduced query. We prove the soundness and completeness of our approach and demonstrate that it outperforms the state-of-the-art on existing benchmarks, as well as a new, more challenging benchmark set from the planning domain.

\end{abstract}

\section{Introduction}\label{sec:introduction}
Boolean Satisfiability (SAT) solvers have been shown to efficiently solve a number of NP-hard problems in areas such as AI planning \cite{kautz1992planning}, verification \cite{clarke2001bounded}, and software testing \cite{cadar2008exe}. Satisfiability Modulo Theories (SMT) solvers are a natural extension to SAT solvers that can reason about first-order structures with background theories~\cite{barrett-smtbookch21}, allowing them to tackle more general problems or to accept more succinct inputs. For example, SMT solvers can reason about bit-vectors \cite{Brummayer2009boolector}, floating-point numbers \cite{rummer2010smt}, strings \cite{bjorner2012smt}, and algebraic data types (ADTs).

The power behind ADTs lies in how they can succinctly express complex structures
at a high-level of abstraction while avoiding common programming pitfalls, like null pointer dereferencing \cite{hoare1975recursive}.
For most of their history, ADTs lived exclusively inside functional programming languages, like NPL \cite{burstall1977design}, Standard ML \cite{milner1997definition}, and Haskell \cite{hudak2007history}.
Recently, however, the interest in ADTs has exploded with a number of mainstream languages being released with support for ADTs, e.g., Rust \cite{jung2021safe}, or have added support, e.g., Python \cite{pythonADT} and Java \cite{javaADTs}. 

Automated reasoning about ADTs is important because this construct appears in many different software applications.
As the popularity of ADTs grows,
the demand for efficient SMT solvers
will continue to increase. Unfortunately, the state-of-the-art tools in this space are already struggling to keep up. We demonstrate this empirically by generating a new benchmark set and showing that existing solvers, working together, are only able to solve 56.2\% of the new queries in under 20 minutes per query (Sec.~\ref{sec:rq2}).

This imbalance between programming languages and SMT solvers is due to a gap in the SMT solving literature. \citeN{Oppen1980} was the first to give a decision procedure for the quantifier-free theory
but ADTs do not seem to have permeated the community much further. In 2003, a consorted effort to unify the SMT community began with the first official input standard and competition, called SMT-LIB and SMTCOMP \cite{barrett2011smt}, respectively. 
ADTs were not officially integrated into the standard until 2017, as part of version 2.6 \cite{smt}. 
In the most recent iteration of SMTCOMP, only two solvers participated in the ADT track, the least of any track, and both solvers use a variation of the same solving approach: a \emph{lazy} SMT architecture combined with theory-specific reasoning based on the work by Oppen in 1980 (see Sec.~\ref{sec:related}).

\begin{figure*}[t]
    \centering
    \begin{subfigure}[t]{0.17\textwidth}
        \centering
        \includegraphics[width=\textwidth]{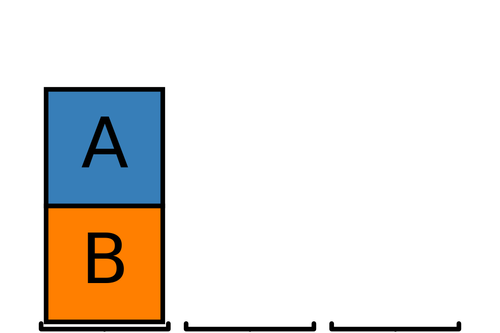}
        \caption{Initial configuration}
        \label{fig:blocksworld-start}
    \end{subfigure}%
    ~ 
    \begin{subfigure}[t]{0.17\textwidth}
        \centering
        \includegraphics[width=\textwidth]{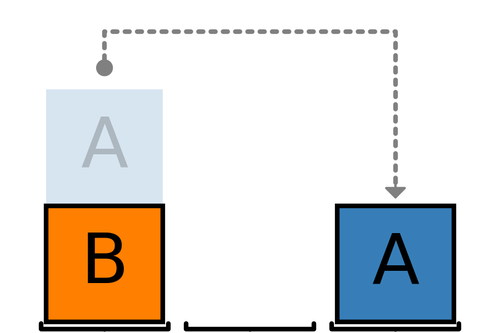}
        \caption{First move}
        \label{fig:blocksworld-step1}
    \end{subfigure}%
    ~ 
    \begin{subfigure}[t]{0.17\textwidth}
        \centering
        \includegraphics[width=\textwidth]{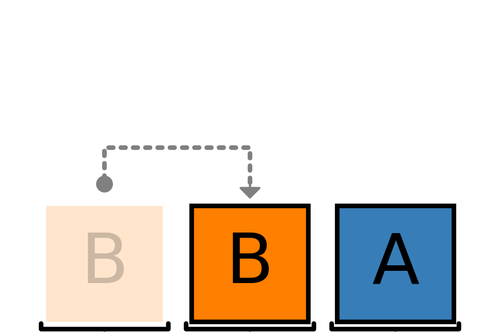}
        \caption{Second move}
        \label{fig:blocksworld-step2}
    \end{subfigure}%
    ~ 
    \begin{subfigure}[t]{0.17\textwidth}
        \centering
        \includegraphics[width=\textwidth]{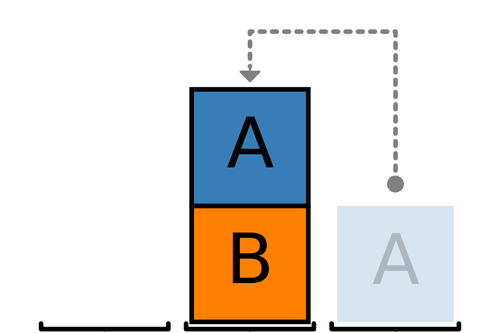}
        \caption{Third move}
        \label{fig:blocksworld-step3}
    \end{subfigure}%
    ~ 
    \begin{subfigure}[t]{0.17\textwidth}
        \centering
        \includegraphics[width=\textwidth]{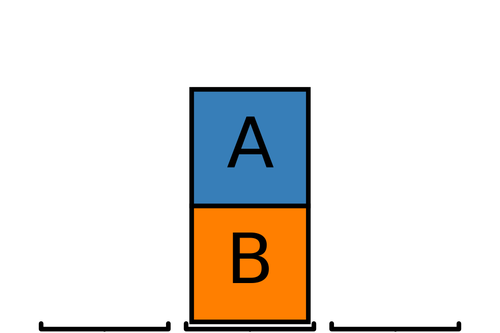}
        \caption{Target configuration}
        \label{fig:blocksworld-end}
    \end{subfigure}

    \caption{Solution (\ref{fig:blocksworld-step1}, \ref{fig:blocksworld-step2}, and \ref{fig:blocksworld-step3}) to a simple blocks world puzzle. \ref{fig:blocksworld-start} is the initial configuration; \ref{fig:blocksworld-end} is the target configuration.}
    \label{fig:blocksworld}
\end{figure*}

In this paper, we propose a new solving technique that departs from the standard approach in the community. Instead of a lazy approach, we take an \emph{eager} approach~\cite{barrett-smtbookch21} that translates the original SMT formula into an equi-satisfiable formula without ADT predicates or functions. Our work fills the gap in the literature on SMT solving for ADT queries, and, by doing so, solves more queries than existing solvers (see Sec.~\ref{sec:rq1}). More importantly, we make the largest empirical contribution to the solving community on SMTCOMP benchmarks, solving different queries than existing tools (see Sec.~\ref{sec:rq2}).

\subsection{Overview and Contributions}
The rest of this paper is organized as follows. In Sec.~\ref{sec:illustrative} we describe ADTs, satisfiability, and our approach through an example planning problem called blocks world. In Sec.~\ref{sec:background} we formally define ADTs and give the necessary background on first-order logic and model theory to understand our approach. In Sec.~\ref{sec:approach} we describe our eager reduction from ADT queries to queries containing only uninterpreted functions (UF). Sec.~\ref{sec:approach} includes a proof of soundness and completeness along with a complexity analysis. In Sec.~\ref{sec:evaluation} we describe a prototype implementation of our approach, called Algaroba, and we evaluate it over two research questions. We find that Algaroba outperforms the state-of-the-art overall and in terms of contribution to the solving community. Sec.~\ref{sec:evaluation} also describes a new benchmark set consisting of blocks world queries. This set fills an important gap in the existing literature: the queries in this set contain all important kinds of ADTs, but are not easy to solve. We survey related work in Sec.~\ref{sec:related} and then conclude in Sec.~\ref{sec:conclusions} with future work. Overall, we make the following contributions.
\begin{enumerate}
    \item We define a notion of query \emph{depth} in Sec.~\ref{subsec:adttheory} and a finite, quantifier-free reduction from ADT queries to UF queries that uses depths. This is a new eager solving approach.
    \item We prove the soundness and completeness of our approach and show that it generates at most a finite number of assertions.
    \item We generate a new benchmark set that contains all important kinds of ADTs but is not trivial to solve. Existing benchmarks do not enjoy both these properties.
    \item We implement our reduction approach in a prototype tool called Algaroba and we compare its performance to the state-of-the-art. We find that Algaroba outperforms existing tools and that it makes the largest empirical contribution to the community of solvers.
\end{enumerate}
\section{Illustrative Example and New Benchmark}\label{sec:illustrative}
To better understand ADTs, satisfiability queries, our approach, and the benchmark set that we generate in Sec.~\ref{sec:evaluation}, consider the classic blocks world problem first proposed by \citeN{winograd1971procedures}. We use a version of the problem that \citeN{sussman1973computational} used to illustrate 
the Sussman anomaly \cite{russell2010artificial} and that \citeN{gupta1992complexity} used to show that the associated decision problem is NP-Hard. 

In the simplified version of the blocks world problem, there is a table with (possibly empty) stacks of blocks on it (an initial configuration), a desired way that the stacks should be arranged (a target configuration), and three rules:
\begin{enumerate}[itemsep=0mm]
    \item blocks can only be taken from the top of a stack;
    \item blocks can only be placed on the top of a stack; and
    \item only one block can be moved at a time. 
\end{enumerate}
The general problem is to find a sequence of legal moves that leads from the initial configuration to the target configuration. The associated decision problem is to determine if there is such a sequence of length less than some input $k$.

Fig.~\ref{fig:blocksworld} shows an example blocks world solution. The initial configuration is given in Fig.~\ref{fig:blocksworld-start}, the target configuration in Fig.~\ref{fig:blocksworld-end}, and Figs.~\ref{fig:blocksworld-step1}, \ref{fig:blocksworld-step2}, and \ref{fig:blocksworld-step3} show a sequence of three legal moves that solve the problem, where faded blocks denote the previous position of the block moved at that step. 

The blocks world problem is a useful illustrative example for an ADTs solver because the encoding uses three important kinds of ADTs: sum, product, and inductive types. The following OCaml code gives the required type definitions for the example in Fig.~\ref{fig:blocksworld}.
\begin{lstlisting}[language=caml,escapechar=\$]
type block = A | B $\label{line:enum}$
type tower = $\label{line:inductive}$
  | Empty
  | Stack of {top: block; rest: tower} 
type config = $\label{line:record}$
  | table of {l:tower; c:tower; r:tower}
\end{lstlisting}
Specifically, this code defines an enumerated type for blocks (``block'' at line \ref{line:enum}), a record type for table configurations (``config'' at line \ref{line:record}), and an inductive type for stacks (``tower'' at line \ref{line:inductive}).
Variables of an enumerated type can take on any of the values listed in the type definition. For example, variables of type ``block'' can take on the values ``A'' or ``B.'' Variables of a record type can take on any combination of values of the type arguments listed in the type definition. For example, variables of type ``config'' can take on a triple of any three ``tower'' values. Enumerated types are the simplest form of a \emph{sum} type, while records are the simplest form of a \emph{product} type.
ADTs allow for definitions that are both sum and product types. For example, variables of type ``tower'' can either be ``Empty'' or they can be a ``Stack'' but not both (sum). When these variables take on a ``Stack'' value, they are a pair of ``block'' and ``tower'' values (product).
Notice that the definition of ``tower'' depends on itself. This makes ``tower`` an \emph{inductive} type as well.

The blocks world problem is a useful illustrative example for satisfiability queries because satisfibility-based solutions for similar planning problems have been around for decades \cite{kautz1996pushing}, and encoding the problem as a satisfiability problem is simple when using bounded model checking \cite{biere1999symbolic}, a standard and well known technique. Specifically, the bounded model checking-based encoding is given by a transition system and a specification. The transition system starts at the initial configuration, and, at each step, makes a non-deterministic choice of which legal move to make. The specification is that the transition system never reaches the target configuration. Given this transition system, specification, and a bound $k$, bounded model checking will generate a satisfiability query whose solutions are assignments to the non-deterministic choices---concrete choices that get us to the target configuration from the initial configuration in less than $k$ steps. SMT solvers, like our new solver, generate these solutions or prove that none exist. We use this encoding in Sec.~\ref{sec:evaluation} to generate a new benchmark set.

The blocks world problem also gives a useful intuition for our approach. While variables of inductive data types, like ``tower,'' can take on arbitrarily large values (e.g., a stack of a million ``A'' blocks), there is a bound on the size of relevant values. For the blocks world problem in Fig.~\ref{fig:blocksworld} it would never make sense to have a tower of size greater than two. Such a bound exists for all quantifier-free ADT queries; the problem is that automatically inferring the bound and using it to speed up solving was non-trivial.
In this paper, we give an automated procedure for computing an over-approximation of this bound, and then we use this over-approximation to replace ADT functions with uninterpreted functions along with quantifier-free axioms.

\section{Background}\label{sec:background}
We assume a basic understanding of first-order logic. For a complete introduction, we refer the reader to \citeN{lubarsky2008ian}.
Many-sorted first-order logic is like first-order logic but with the universe partitioned into different sorts \cite{barrett-smtbookch21}. 
We use many-sorted first-order logic and assume all terms are well-sorted, i.e. that we never apply a function to a term of the incorrect sort. In practice, standard type checking algorithms will catch these issues.

A first-order theory is a set of formulas (axioms). For example Equality and Uninterpreted Functions \cite{burch1994automatic} (\UF) is the theory which interprets the symbol ``$=$'' using the usual axioms of equality (reflexivity, symmetry, and transitivity). Furthermore, every function must satisfy congruence: applications to equal inputs give equal outputs. 

Satisfiability Modulo Theories (SMT) solvers~\cite{barrett-smtbookch21} take a first-order theory and a formula, and return \textbf{sat} if there is an assignment to all functions and variables that satisfies the formula and the theory axioms.
More formally, a \emph{structure} is a universe along with a function that describes how all non-logical symbols must be interpreted over the universe. When a structure $\M$ satisfies a formula $\phi$, then we say that $\M$ is a \emph{model} of $\phi$ and we denote this as $\M \models \phi$. 
As a slight abuse of notation, for a set of formulas $\Phi = \{\phi_i\}$, we use $\M \models \Phi$ to mean $\M \models \bigwedge_1^n \phi_i$. 
Similarly, we use $\phi \models \psi$ to mean every model of $\phi$ is a model of $\psi$.
For a model $\M$, we use the notation $\M[x]$ or $\M[f]$ to represent the variable $x$ or the function $f$ interpreted in the model $\M$.

An \emph{SMT query} is a formula-theory pair. For a formula $\phi$, there are free variables (which we will just call variables)
and uninterpreted function symbols
that we are trying to find an interpretation for. 
We say that an SMT query ($\phi, \textbf{T}$) is \textbf{sat} iff there exists a model $\M$ such that $\M \models T \cup \{\phi\}$. Otherwise we say the query is \textbf{unsat}.

\subsection{Theory of Algebraic Data Types} \label{subsec:adttheory}
We denote the theory of algebraic data types as \DT{}. It contains the full theory of \textbf{UF} and additional structure given by:

\begin{defi}[\DT{} \cite{smt}] \label{def:adt}
    An ADT $\A$ with sort $\sigma$ is a tuple consisting of:    
    \begin{itemize}
        \item A finite set of constructor functions $\A^C $, where we say a function $f: \sigma_1 \times ... \times \sigma_l\rightarrow \sigma$ has sort $\sigma$ and arity $l$
        \item A finite set of selectors $\A^S$, such that there are $l$ selectors $f^1, ..., f^l$ for each constructor $f \in \A^C$ with arity $l$.
        \item A finite set of testers $\A^T$ and a bijection $p: \A^C \rightarrow \A^T$ which sends $f \mapsto is_f$ where $is_f: \sigma \rightarrow \{\texttt{True}, \texttt{False}\}$
    \end{itemize}
    Every ADT $\A$ satisfies that axioms given in Fig.~\ref{fig:axiomsofadt}, where we say that $s$ is a \emph{child} of $t$ if it can be obtained by applying a sequence of selectors to $t$.
\end{defi}

\begin{figure}
    \begin{tcolorbox}
        \begin{itemize}
            \item $\forall \, \vec{s} \; is\mhyphen f (f (\vec{s})) = \texttt{True}$
            \item  $\forall \, \vec{r} \; is\mhyphen f (g(\vec{r})) = \texttt{False}$ for constructors $g \neq f$
            \item $\forall \, \vec{t} \; f^i (f (\vec{s})) = \vec{s}_i$ for every selector $f^i$ of $f$
            \item $\forall \, t \; is\mhyphen f(t) \rightarrow \exists \, \vec{s} \; f(\vec{s}) = t$ %
            \item $\forall \, t \, \forall \, s \;$ if $s$ is a child of $t$, then $s \neq t$
        \end{itemize}
    \end{tcolorbox}
    \caption{Axioms of an ADT for every constructor $f$, corresponding tester ($is\mhyphen f$), and corresponding selectors ($f^i$).}
    \label{fig:axiomsofadt}
\end{figure}

As an example, the ``tower'' definition from earlier
uses two constructors: ``Empty'' and ``Stack.'' These are the two possible ways to build a ``tower.'' ``Empty'' is a function that takes no inputs and outputs a ``tower.'' ``Stack'' is a function that takes a ``block'' and a ``tower'' and outputs a ``tower.'' 
Each corresponding constructor has a set of selectors. ``Empty'' has no selectors and so we call it a \emph{constant}, but ``Stack'' has two selectors, ``top'' and ``rest.'' In OCaml, we apply selectors using ``dot'' notation, e.g., ``x.top.'' Selectors can be thought of as de-constructors---we use them to get back the terms constructed a ``tower.''
The ``tower'' definition implicitly defines two testers: ``is-Empty'' and ``is-Stack.'' These predicates take a ``tower'' and return true iff the argument was built using the matching constructor.

An ADT term is any expression of an ADT sort. The set of \emph{normal} ADT terms is the smallest set containing
\begin{enumerate}[itemsep=0mm]
    \item constants (0-ary constructors),
    \item constructors applied to only normal ADT terms.
\end{enumerate}
It is useful to think of normal terms as trees: the constants are leaves and we can build larger trees by applying constructors to sub-trees. These trees can be deconstructed by applying selectors. Our approach depends on \emph{depth}: a measure of how far we deconstruct these trees.

If we apply a sequence of selectors $f_1^{n_1}, ..., f_l^{n_l}$ to a term $t$: $f_l^{n_l}(...(f_1^{n_1}(t)...)$, then say we are selecting up to a \emph{depth} $l$. We use depth values to transform \DT{} queries into \UF{} queries through a \emph{theory reduction}. 

\begin{defi}[Theory Reduction] 
\label{def:reduction}
        A theory $T$ \textbf{reduces} to a theory $R$ if there is a computable $m$ such that $$(\psi, T) \text{ is \textbf{sat}} \leftrightarrow (m(\psi), R) \text{ is \textbf{sat}}$$
\end{defi}

An \emph{theory literal} is a component of logical formula with no conjunctions ($\land$) or disjunctions ($\lor$). These are the base units of SMT solving and are the equivalent of \emph{literals} in SAT solving. Our approach is easier to understand when queries are \emph{flat}. A formula $\phi$ is flat if the only theory literals are of the form $\neg (x_1 = x_2)$, $x_1 = x_2$, $x_1 = g(x_2, ..., x_n)$ where $x_i$ are variables. Any query can be ``flattened'' so for simplicity, we assume that every input query $\psi$ is flat.

\section{Proposed Approach: Eager ADT Solving}\label{sec:approach}
Our solver takes a quantifier-free formula $\psi$ in \DT{} and returns \textbf{sat} or \textbf{unsat}. It does this by reducing to a quantifier-free formula $\psi^*$ in \UF{} and then applies a \UF{} solver to get a result. 
Our reduction encodes the axioms of \DT{} into quantifier-free formulas over \UF{}. We cannot do this directly, since \DT{} axioms have universal quantifiers. Solving theories with universal quantifiers is expensive and is not supported by many SMT solvers. Instead, 
we will only instantiate our axioms over terms that are relevant to the query. If our universe is finite, we deal with this by instantiating the entire universe (see Section \ref{subsec:finiteuniverse}).

Our reduction is given in Fig.~\ref{fig:reduction}. Rules A and B correspond to rewriting constructor and selector applications respectively so that they work well with other constructor, selectors and testers. We add in additional axioms 1 and 2 to ensure that each term is only satisfied by one tester and it is satisfied by a tester corresponding to a constant iff it is a constant. Axiom 3 encodes acyclicality. Axiom 2 contains existential quantifiers, but these are handled by \emph{Skolemization}, i.e. replacing variables bound by an existential quantifier with free variables.

\begin{figure}[t]
    \centering
    \includegraphics[width=0.6\linewidth]{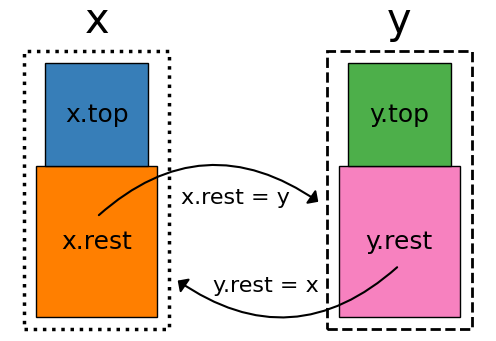}
    \caption{Visual representation of an \textbf{unsat} query.}
    \label{fig:tower-cycle}
\end{figure}

To better understand axiom 3 and acyclicality, consider the following example query. Let \texttt{x} and \texttt{y} be of type \texttt{tower} defined in Sec.~\ref{sec:illustrative}. The query
\begin{lstlisting}
(is-Stack x) && (is-Stack y) 
  && (y = x.rest) && (x = y.rest)
\end{lstlisting}
is \textbf{unsat} because any satisfying assignment would need to violate acyclicality. Fig.~\ref{fig:tower-cycle} illustrates this. There is a circular dependency between \texttt{x} and \texttt{y} that cannot be resolved in the \DT{} theory. To avoid spurious models, we must encode this acyclicality property in our reduction. However, the problem is that this is an ``infinite'' property that we have to instantiate in a finite number of quantifier-free queries.

\begin{figure*}[t!]
\begin{minipage}[t]{0.5\linewidth}
\centering
\begin{tcolorbox}
    \begin{enumerate}[label =  \Alph*.]
        \item $f(\vec{s}) = t \Longrightarrow f(\vec{s}) = t \land is \mhyphen f(t) \land \bigwedge_{i=1}^l f^i(t) = \vec{s}_i$
        \item $f^j(t) = t_j \Longrightarrow f^j(t) = t_j \; \land $
        \setlength\topsep{0pt}
        \setlength\parskip{0pt}
        \begin{flushright}
        $[\textit{is-f} (t) \rightarrow [\exists \vec{s} [f(\vec{s}) = t \land \bigwedge_{i=1}^l f^i(t) = \vec{s}_i]]]$
        \end{flushright}
    \end{enumerate}
    
\end{tcolorbox}
\subcaption{Rewrite every term in $\psi$ using rules A and B to get $\Tilde{\psi}$.}
\label{subfig:reductionrules}
\end{minipage}
\begin{minipage}[t]{0.5\linewidth}
\centering
\begin{tcolorbox}
\begin{enumerate}[label = \arabic*.]
    \item Add $\bigvee_{i=1}^{|\A_T|} [is \mhyphen f_i(t) \land\bigwedge_{j = 1, j \neq i}^{|\A_T|} \neg is \mhyphen f_j(t)]$
    \item For any constant constructor $c$, add $is \mhyphen c (t) \leftrightarrow c = t$
    \item If $s$ is a child of t with depth $< k$, add $s \neq t$
\end{enumerate}

\end{tcolorbox}

\subcaption{
Instantiate axioms 1, 2, and 3 for all variables $t$ to get $\{\phi_i\}$. 
}
\label{subfig:reductionaxioms}
\end{minipage}
\caption{Rewrite rules (a) and additional axioms (b) to reduce of a flat formula $\psi$ in \DT{} to $\psi^* = \Tilde{\psi} \land \phi_1 \land ... \land \phi_m$ in \UF{}}\label{fig:axiomandreduction}
\label{fig:reduction}
\end{figure*}

A key insight of this work is a sufficient condition that captures the acyclicality of a query. Specifically, let $V = \{x_1, ..., x_k\}$ be the set of variables in the input query after flattening and Skolemization. We only need to enforce acyclicality up to a depth $k$. 
This enforcement is accomplished by axiom 3 in Fig (\ref{subfig:reductionrules}).

More specifically, axiom 3 will say for every $x \in V$, for every (well-typed) sequence of selectors $f_1^{n_1}, ..., f_l^{n_l}$ for $l \leq k$, we add the following axiom:
\begin{align*}
    (\textit{is-f}_1(t) &\land ... \land \textit{is-f}_l (f_{l-1}^{n_{l-1}}(...(f_1^{n_1}(t)...)) \\
    &\rightarrow f_l^{n_l}(f_{l-1}^{n_{l-1}}( ... (f_1^{n_1} (t) ...) \neq t
\end{align*}
Here the antecedent serves to guard the selector applications, i.e. to ensure that we apply selectors to appropriately constructed term, since incorrectly applied selectors have unspecified behavior and so we do not need to worry about  them creating a cycle.

In practice, we do not use the same $k$ for all ADTs, rather for each $\A$, we compute a $k_\A$ which is the number of variables of that ADT added to the number of variables of any other ADT that refers to $\A$ and has $\A$ refer to it. This is easily computed by storing the relative relationship of all ADTs in a graph and doing a depth-first search.

\subsection{Finite Universe Instantiation} \label{subsec:finiteuniverse}

If we recognize an ADT has a finite universe, we \emph{instantiate the universe} in $\psi^*$, i.e. we create constants for every term in the universe. This is a source of double exponential blowup, but ADTs with finite universes are rare and often small enough to prevent noticeable blowup.

One adversarial case where finite universes could occur, is if there is an ADT that is records of records of enums:
\begin{lstlisting}[language=caml,escapechar=\$]
type enum = A | B 
type rec1 = j of {l: enum; r: enum}
type rec2 = k of {m: rec1; s: rec1}
\end{lstlisting}
Here, \texttt{enum} has a universe of size two, \texttt{rec1} has a universe of size four, and \texttt{rec2} has a universe of size $16$. This gives a double exponential blowup since we are creating variables to represent each term of each data type. 

\subsection{Proof of Correctness} \label{subsec:proof}
In this proof when we refer to a rule or axiom, we mean those from Fig.~\ref{subfig:reductionrules} and Fig.~\ref{subfig:reductionaxioms}, respectively.

\begin{theorem}\label{thm:numvarsreduction}
    Say $\psi$ is a flattened ADT-formula. Then $(\psi, \DT{})$ is \textbf{sat} $\leftrightarrow (\psi^*, \UF{})$ is \textbf{sat}, where we compute $\psi^*$ from $\psi$ using Fig.~\ref{fig:reduction} rules A, B and axioms 1, 2, 3.
\end{theorem}
\begin{proof}
    \underline{$\rightarrow$:} If $(\psi, \DT{})$ is \textbf{sat}, then there is a model $\M \models \psi \cup \DT{}$. $\psi^*$ is $\psi$ modified according to rules A and B and axioms 1, 2, and 3. Each of these rules and axioms is consistent with the axioms of \DT{} in Definition \ref{def:adt} and thus $\M \models \DT{} \cup \psi^*$. Since every model in \DT{} is a model in \UF{}, $\M \models \UF{} \cup \psi^*$. Thus, $(\psi^*, \UF{})$ is \textbf{sat}
        
    \underline{$\leftarrow$:} If  $(\psi^*, \UF{})$ is \textbf{sat}, then we know that there is some model $\N \models \UF{} \cup \psi^*$. We want to modify $\N$ to create a model $\M \models \DT{} \cup \psi$. As we describe in Section \ref{subsec:finiteuniverse}, if the universe is finite, we manually instantiate all of the ADT normal terms in the query. Thus, in the finite universe case, it must be that $\DT{} \models \psi$. From here on we will assume the \DT{} universe is infinite

   Let $\beta^*$ be a satisfying assignment to the propositional structure of $\psi^*$, i.e. a minimal conjunction of theory literals such that $\N \models \beta^*$ and $\beta^* \models\psi^*$. There must be some largest subset of $\beta^*$, call it $\beta$, such that all of the theory literals in $\beta$ appear in $\psi$.
    
    The modification from $\psi$ to $\psi^*$ involves either replacing a theory literal with a conjunction of theory literals or tacking on conjunctions of theory literals to the end of the formula. Since the theory literals we remove from $\beta^*$ will also be removed from $\psi^*$, it must be that $\beta \models\psi$.

    Since we want $\M \models \DT{}$, we set the universe of $\M$ to be all of the ADT normal terms. Consider the set of variables that appear in $\beta^*$ which we call $V = \{x_1, ..., x_k\}$. We describe an algorithm that for all $x \in V$, sets $\M[x]$ to some ADT normal term, such that we ultimately get $\M \models \beta$.

    $V$ is the set of all variables that appear in $\psi^*$, thus for each $x \in V$, by axiom 1 there is exactly one tester $\textit{is-f}$ such that $\N \models \textit{is-f}(x)$.

    There are two  ``base cases'' for our construction of $\M$. First, if $f$ is some constant constructor, by axiom 2 of the reduction, we know that $\N[x] = \N[f]$, so we set $\M[x] \triangleq \M[f]$. Second, if $x$ is some variable that is never set equal to some constructor application or selected from (either directly or transitively) then we set $\M[x]$ to an ADT normal term. Since our \DT{} universe is infinite, we will specifically pick an ADT normal term $t$ such that it takes at least $k+1$ selector applications to get to any of the ADT normal terms that we have already set. This will prevent any of our different ADT normal term assignments from interfering with each other---they are too far away from each other in the infinite universe.

    If we are not in one of these base cases, we know that $f$ is an $l$-ary constructor for some $l > 0$. Since $x$ was either constructed or selected, there are variables $y_1, ..., y_l$ in $V$ such that $\N \models \bigwedge_{i=1}^l f^i(x) = y_i$. 
    Note that these variables are from the original query if $x$ is equal to a constructor application, or from Skolemization if $x$ is selected from. 

    Continuing our construction of $\M$, we recurse on these $y_i$ that have not already been assigned in $\M$. We will eventually hit a base case, since there are a finite number of selector/constructor applications in our original query.
    For each $i$, we set $\M[f^i](\M[x]) \triangleq \M[y_i]$.
    Finally, we set 
    $\M[x] \triangleq \M[f](\M[y_1], ..., \M[y_l])$.
    
    If it were possible to have
    \begin{equation} \label{eq:badconstructor}
        \beta \models f(y_1, ..., y_l) \neq x
    \end{equation}
    hold, then we would have $\M \not\models \beta$ and our current proof attempt would not go through. However, we will show that this is never the case.
    
    Note that since $\N \models \beta^*$, if $\beta^*$ asserts anything about selector applications, these selector applications must be consistent with $\N$. Also, $\beta^*$ must assert something about selector applications, since we know that $x$ is either equal to a constructor application or is selected from in the query. Thus, $\beta^* \models \bigwedge_{i=1}^l f^i(x) = y_i$, meaning that $\beta^*$ asserts the correct selector behavior. We now use this to guarantee the correct constructor behavior.
    
    There are two ways that incorrect constructor behavior could occur in $\beta^*$:
    \begin{itemize}
        \item If $\beta \models f(y_1, ..., y_l) = x$ which contradicts Equation (\ref{eq:badconstructor}).
        \item If $\beta \models \bigwedge_{i=1}^l y_i = f^i(x)$, but then by rule B, we would still have $\beta^* \models f(y_1, ..., y_l) = x$ which also contradicts Equation (\ref{eq:badconstructor}) since $\beta^* \models \beta$
    \end{itemize}

    We iterate this construction of $\M$ for each variable in $V$ for at most $k$ rounds, since there are $k$ total variables in $V$. Thus, since $\psi^*$  has the acyclicality axiom 3 instantiated up to a depth $k$, we do not create any cycles in $\M$
    
    We can also see that $\M \models \beta$ since each theory literal $\beta_i$ in $\beta = \bigwedge_{i=1}^p \beta_i$ will be an equality $x = y$, disequality $x \neq y$, selector application $f^j(y) = x$, a tester application $\textit{is-f} (x)$, or a constructor application $f(x_1, ..., x_l) = y$. If it is any of these, then $\M \models \beta_i$ by how we defined $\M$. Note that if it was a constructor application, then by rule A $\beta^*$ would have the respective selector applications and thus our construction of $\M$ would satisfy $\beta_i$.

    Thus, $\M  \models \DT{} \cup \psi$ and so $(\psi, \DT{})$ is \textbf{sat}.
    
\end{proof}

\subsection{Complexity Analysis} \label{subsec:complexity}
This reduction can create an exponential blow up in the size of the query. We know the depth $k$ is at most linear in the size of the query, since it is the number of variables (and flattening our query will only create a linear blowup).
However, for a term $x$ of the \texttt{tree} type definition below, the number of selector applications goes up to depth $k$ is $2^{k+1} - 2$, thus giving us an exponential blowup in the number of terms.
\begin{lstlisting}
type tree = 
  | Leaf
  | Node of {left: tree; right: tree} 
\end{lstlisting}

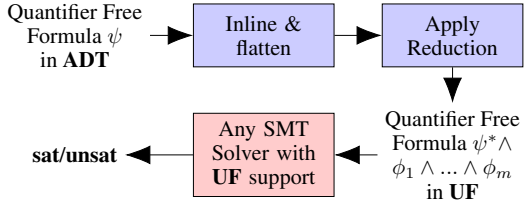
\begin{figure}[t] 

\begin{center}
\resizebox{0.85\linewidth}{!}{
\begin{tikzpicture}[node distance=2cm]

\node (start) [align=center] {Quantifier Free \\ Formula $\psi$ \\ in \textbf{ADT}};
\node (in1) [rectangle, draw, fill=blue!20, text width=2cm, text centered, minimum height=1cm, right of=start, xshift=1cm] {Inline \& flatten};
\node (in2) [rectangle, draw, fill=blue!20, text width=2cm, text centered, minimum height=1cm, right of=in1, xshift=1cm] {Apply Reduction};
\node (end) [align=center, below of=in2, yshift= 0.1 cm] { Quantifier Free \\ Formula $\psi^* \land$ \\ $\phi_1 \land ... \land \phi_m$ \\ in \textbf{UF}};
\node (solve) [rectangle, draw, fill=red!20, text width=2cm, text centered, minimum height=1cm, left of=end, xshift= -1cm] {Any SMT Solver with \textbf{UF} support};
\node (result) [align=center, left of=solve, xshift= -1cm] {\textbf{sat}/\textbf{unsat}};

\draw [-{Latex[length=4mm]}] (start) -- (in1);
\draw [-{Latex[length=4mm]}] (in1) -- (in2);
\draw [-{Latex[length=4mm]}] (in2) -- (end);
\draw [-{Latex[length=4mm]}] (end) -- (solve);
\draw [-{Latex[length=4mm]}] (solve) -- (result);
\end{tikzpicture}
}
    
\end{center}
\caption{Reduction architecture. 
}

\label{fig:reduction}
\end{figure}

\section{Empirical Evaluation}\label{sec:evaluation}
\begin{figure*}[t]
    \centering
    \begin{subfigure}[t]{0.5\textwidth}
        \centering
        \includegraphics[width=0.8\linewidth]{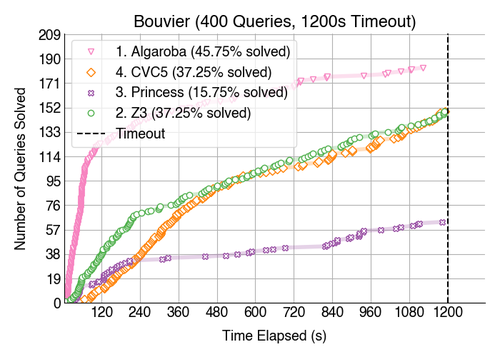}
        \caption{Bouvier benchmark set.}
        \label{fig:bouvier-line}
    \end{subfigure}%
    ~ 
    \begin{subfigure}[t]{0.5\textwidth}
        \centering
        \includegraphics[width=0.8\linewidth]{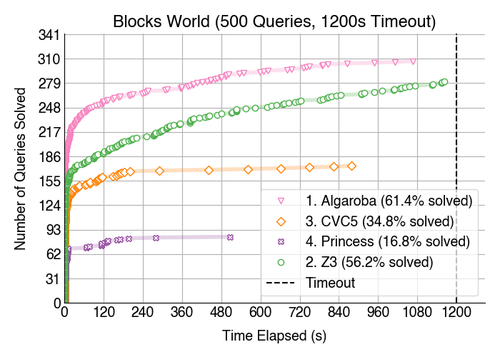}
        \caption{Blocks world benchmark set.}
        \label{fig:blocksworld-line}
    \end{subfigure}
    \caption{Number of queries solved ($y$) in less than $x$ seconds for Bouvier and blocks world benchmark sets using a 1200s timeout. Higher (more queries solved) left (in less time) points are better. The legend lists the contribution rank and percentage of queries solved for each solver. Algaroba solves the most queries and achieves the highest contribution rank for both sets.}
    \label{fig:eval}
\end{figure*}
In this section we empirically compare the performance of our approach to  state-of-the-art solvers. Specifically, we aim to answer the following research questions.
\begin{enumerate}[label=RQ\arabic*,leftmargin=*]
    \item How does the overall performance of our approach compare to the state-of-the-art?
    \item How complementary is the performance of our approach to that of existing solvers?
\end{enumerate}

We implement a prototype of our approach in approximately 2900 lines of OCaml code using the Z3 API. We call this prototype Algaroba and show a high-level architecture diagram in Fig.~\ref{fig:reduction}. Algaroba includes a number of simple optimizations, like hash-consing \cite{hash-consing}, incremental solving, and theory-specific query simplifications. All experiments are conducted on an Ubuntu workstation with nine Intel(R) Core(TM) i9-9900X CPUs running at 3.50GHz and with 62 GB of RAM. All solvers were given a 1200 second timeout on each query to be consistent with SMTCOMP. The state-of-the-art solvers in this space are CVC5 (we use version 1.0.6-dev.214.97a64fc16) and Z3 (we use version 4.12.2). We also include Princess (latest release as of 2023-06-19) in our evaluation since it is the most related approach. We describe all three solvers in Sec.~\ref{sec:related}.

Our evaluation covers two existing benchmark sets from SMTCOMP, one originally from \citeN{bouvier2021vlsat} and one originally from \citeN{barrett2007abstract}. 
These two benchmark sets are useful but limited: every solver succeeds on every query from Barrett et al. so it is difficult to draw performance conclusions; Bouvier queries are more challenging but only contain sum types. 

To address these limitations, we introduce a new benchmark set consisting of randomly generated blocks world queries. Blocks world queries, which we describe in Sec.~\ref{sec:illustrative}, are more challenging to solve than those from Barrett et al. and, unlike those from Bouvier, contain sum, product, and inductive types. To generate blocks world queries we use the same table configuration as in Sec.~\ref{sec:illustrative} (three places for towers), but we randomly select a set of blocks (ranging between two to 26) and we randomly generate an initial and target configuration (two sets of three random block towers). We call these three random samples a blocks world setup. For each blocks world setup, we randomly sample a set of step numbers (ranging from one to two times the number of blocks) and generate a blocks world query for each step number. This process resulted in 500 individual queries that each ask ``can we get from this initial configuration to this target configuration in exactly this number of steps?''

\subsection{RQ1: Overall Performance}\label{sec:rq1}
To answer our first research question, we time the execution of Algaroba, CVC5, Princess, and Z3 on all queries in all three benchmark sets. When more than one solver terminates on a given query we compare the results to check for disagreements. There was not a single case where one solver returned SAT and another returned UNSAT therefore we focus the remainder of our evaluation on execution times.

For the Barrett et al. benchmark set, which consists of 8000 queries, every solver successfully terminates on every query within the timeout. CVC5 performs the best on average with an average solve time of 0.05 seconds (compared to 0.08 seconds for Algaroba and 0.10 seconds for Z3). Z3 performed the most consistently with a standard deviation of 0.05 seconds (compared to 0.10 seconds for Algaroba and 0.15 seconds for CVC5). Given the magnitude of these values, we conclude that the performance differences between Algaroba, CVC5, and Z3 are negligible on this set. Princess is the slowest (2.20 seconds on average) and least consistent (standard deviation of 1.69 seconds) but still effective.

Results are more interesting for the remaining benchmark sets. Fig.~\ref{fig:bouvier-line} shows the execution times for every solver on every query in the Bouvier benchmark set (excluding timeouts). No solver succeeds on more than half the queries in the set but Algaroba clearly outperforms the rest. In terms of number of queries solved, Algaroba succeeds on 8.5 percentage points more queries than the next best (45.75\% versus CVC5 and Z3's 37.25\%). In terms of average run time on successful queries, Algaroba is 2.30 times faster than the next best (190.11 seconds versus Z3's 437.18 seconds). In terms of standard deviation on successful queries, Algaroba is 1.30 times more consistent than the next best (267.13 seconds versus CVC5's 346.83 seconds).

Fig.~\ref{fig:blocksworld-line} shows the execution times for every solver on every query in the blocks world benchmark set (excluding timeouts). 
Again, Algaroba outperforms the state-of-the-art solvers. Algaroba solves 5.2 percentage points more queries than the next best (61.4\% versus Z3's 56.2\%) but the average and standard deviation results are more complicated. Princess is the fastest and most consistent solver on solved queries (31.89 seconds and 74.77 seconds, respectively) but it succeeds on 44.6 percentage points fewer queries than Algaroba. Compared to Z3, which solves the second most number of queries, Algaroba is 2.00 times faster (87.34 seconds on average compared to Z3's 175.05 seconds) and 1.51 times more consistent in terms of standard deviation (198.67 seconds versus Z3's 300.15 seconds).

Given the overall success of Algaroba on both interesting benchmark sets, we answer RQ1 by concluding that our performance compares favorably to the state-of-the-art.

\subsection{RQ2: Contribution Rank}\label{sec:rq2}
Measuring overall performance is useful but it does not give an accurate perspective on how the community uses these tools. When faced with an SMT query, practitioners are likely to try multiple different solvers until they find one that works. Some may even run all disposable tools in parallel hoping that at least one will terminate \cite{rungta2022billion}. \emph{Contribution ranks} capture this practical perspective by evaluating solvers in terms of how complementary they are to all other solvers in the space. A higher rank means that the solver is contributing more to the community of solvers. 

To evaluate how complementary our approach is to existing solvers, we use SMTCOMP's contribution ranking. This ranking uses the notion of a \emph{virtual best} solver, which is defined as $vb(q, S) \triangleq s(q)$, where $S$ is a set of solvers and $s$ is the solver in $S$ that terminates most quickly on $q$. Informally, the ranking answers, ``which solver can I remove from the virtual best solver to hurt performance the most?''

In terms of number of queries solved (the primary SMTCOMP metric), there is a four-way tie on the Barrett et al. benchmark set---all solvers solve all queries. For both other benchmark sets, Algaroba is ranked highest. For blocks world, the virtual best solver without Algaroba succeeds on 56.2\% of the queries, less than Algaroba on its own. With Algaroba, the virtual best solver succeeds on 62.2\%. For Bouvier, without Algaroba, the virtual best solver succeeds on 64.25\% of the queries. With Algaroba, this number rises to 83.75\%. 
These positive results are in part because Algaroba solves the most queries, but are mainly due to the uniqueness of our approach. CVC5 and Z3 use a similar underlying algorithm, so removing one does not affect the performance of the virtual best solver. On the other hand, while Princess is the most similar approach to our own, their reduction is different enough to not interfere with our ranking. We solve many queries that no other solver can.

Given the winning contribution rank of Algaroba on both interesting benchmark sets, we answer RQ2 by concluding that our performance is complementary to existing solvers -- we solve many benchmarks that no other solver can.

\section{Related Work}\label{sec:related}
Most solvers for quantifier-free ADT queries use a lazy SMT architecture, i.e., they use a theory specific solver to handle the data types and a core solver to handle the logical formula \cite{Sebastiani2007lazy}. A common theory solver will use a combination of congruence closure, syntactic unification, and acyclicality checks \cite{barrett2007abstract, Oppen1980, Reynolds2017codatatypes, Reynolds2018shared}. This is the case for popular SMT solvers like CVC5~\cite{cvc5}, SMTInterpol~\cite{smtinterpol}, and Z3~\cite{z3}. CVC5 and SMTInterpol were the only two participants in the most recent SMTCOMP for quantifier-free ADT queries. We differ in that we take an eager approach.

Princess~\cite{Hojjat} also takes an eager approach. However, Princess reduces queries to \UF{} and Linear Integer Arithmetic (\LIA{}).
\LIA{} makes keeping track of the depth of ADT terms easy, but their reduction results in queries that are more difficult to solve (see Sec.~\ref{sec:evaluation}).

The scope of our work is quantifier-free ADT queries. However, there is existing related work that deals with quantifiers. \citeN{DeAngelis2020chc} and \citeN{Kostyukov} provide approaches to solving ADT Constrained Horn Clauses (CHCs). Other approaches \cite{suter2010abstractions, pham2014unrolling} support restricted forms of recursive functions (called catamorphisms) via partially evaluating these functions. \citeN{kovacs2017terms} provides two decision procedures for quantified ADTs.

\section{Conclusions}\label{sec:conclusions}
As the popularity of ADTs continues to grow,
the demand for efficient SMT solvers that can handle ADTs will increase. Unfortunately, there are few existing solvers in this space and the performance of these solvers can be improved. 

We introduced a reduction from quantifier-free \textbf{ADT} queries to quantifier-free \textbf{UF} queries. This approach is sound, complete, and eager, while most existing approaches are lazy. We implemented a prototype tool of our approach, called Algaroba, and compared it against existing state-of-the-art solvers. We found that, on existing and new benchmarks, we can solve more queries using less time. More importantly, we found that we make the largest empirical contribution to the solving community. 

In the future, we intend to support proof generation, quantifiers, and hybrid eager and lazy approaches. We will also experiment with different back-end solvers and techniques for automatically selecting back-ends per input query.

\bibliography{main}

\begin{thebibliography}{41}
\providecommand{\natexlab}[1]{#1}

\bibitem[{Barbosa et~al.(2022)Barbosa, Barrett, Brain, Kremer, Lachnitt, Mann,
  Mohamed, Mohamed, Niemetz, N{\"o}tzli, Ozdemir, Preiner, Reynolds, Sheng,
  Tinelli, and Zohar}]{cvc5}
Barbosa, H.; Barrett, C.; Brain, M.; Kremer, G.; Lachnitt, H.; Mann, M.;
  Mohamed, A.; Mohamed, M.; Niemetz, A.; N{\"o}tzli, A.; Ozdemir, A.; Preiner,
  M.; Reynolds, A.; Sheng, Y.; Tinelli, C.; and Zohar, Y. 2022.
\newblock cvc5: A Versatile and Industrial-Strength SMT Solver.
\newblock In Fisman, D.; and Rosu, G., eds., \emph{Tools and Algorithms for the
  Construction and Analysis of Systems}, 415--442. Cham: Springer International
  Publishing.
\newblock ISBN 978-3-030-99524-9.

\bibitem[{Barrett et~al.(2011)Barrett, de~Moura, Ranise, Stump, and
  Tinelli}]{barrett2011smt}
Barrett, C.; de~Moura, L.; Ranise, S.; Stump, A.; and Tinelli, C. 2011.
\newblock The SMT-LIB Initiative and the Rise of SMT: (HVC 2010 Award Talk).
\newblock In \emph{Hardware and Software: Verification and Testing: 6th
  International Haifa Verification Conference, HVC 2010, Haifa, Israel, October
  4-7, 2010. Revised Selected Papers 6}, 3--3. Springer.

\bibitem[{Barrett, Fontaine, and Tineli(2017)}]{smt}
Barrett, C.; Fontaine, P.; and Tineli, C. 2017.
\newblock The SMT-LIB Standard Version 2.6.
\newblock
  \url{https://smtlib.cs.uiowa.edu/papers/smt-lib-reference-v2.6-r2017-07-18.pdf}.

\bibitem[{Barrett et~al.(2021)Barrett, Sebastiani, Seshia, and
  Tinelli}]{barrett-smtbookch21}
Barrett, C.; Sebastiani, R.; Seshia, S.~A.; and Tinelli, C. 2021.
\newblock Satisfiability Modulo Theories.
\newblock In Biere, A.; Heule, M.; van Maaren, H.; and Walsh, T., eds.,
  \emph{Handbook of Satisfiability}, chapter~33, 1267--1329. IOS Press, second
  edition.

\bibitem[{Barrett, Shikanian, and Tinelli(2007)}]{barrett2007abstract}
Barrett, C.; Shikanian, I.; and Tinelli, C. 2007.
\newblock An Abstract Decision Procedure for Satisfiability in the Theory of
  Recursive Data Types.
\newblock \emph{Electronic Notes in Theoretical Computer Science}, 174(8):
  23--37.
\newblock Combined Proceedings of the Fourth Workshop on Pragmatics of Decision
  Procedures in Automated Reasoning (PDPAR 2006) and the First International
  Workshop on Probabilistic Automata and Logics (PaUL 2006).

\bibitem[{Biere et~al.(1999)Biere, Cimatti, Clarke, and
  Zhu}]{biere1999symbolic}
Biere, A.; Cimatti, A.; Clarke, E.; and Zhu, Y. 1999.
\newblock Symbolic model checking without BDDs.
\newblock In \emph{Tools and Algorithms for the Construction and Analysis of
  Systems: 5th International Conference, TACAS’99 Held as Part of the Joint
  European Conferences on Theory and Practice of Software, ETAPS’99
  Amsterdam, The Netherlands, March 22--28, 1999 Proceedings 5}, 193--207.
  Springer.

\bibitem[{Bj{\o}rner et~al.(2012)Bj{\o}rner, Ganesh, Michel, and
  Veanes}]{bjorner2012smt}
Bj{\o}rner, N.; Ganesh, V.; Michel, R.; and Veanes, M. 2012.
\newblock An SMT-LIB format for sequences and regular expressions.
\newblock \emph{SMT}, 12: 76--86.

\bibitem[{Bouvier(2021)}]{bouvier2021vlsat}
Bouvier, P. 2021.
\newblock The VLSAT-3 Benchmark Suite.
\newblock \emph{INRIA Technical Report 516}.

\bibitem[{Brummayer and Biere(2009)}]{Brummayer2009boolector}
Brummayer, R.; and Biere, A. 2009.
\newblock Boolector: An Efficient SMT Solver for Bit-Vectors and Arrays.
\newblock In Kowalewski, S.; and Philippou, A., eds., \emph{Tools and
  Algorithms for the Construction and Analysis of Systems}, 174--177. Berlin,
  Heidelberg: Springer Berlin Heidelberg.
\newblock ISBN 978-3-642-00768-2.

\bibitem[{Burch and Dill(1994)}]{burch1994automatic}
Burch, J.~R.; and Dill, D.~L. 1994.
\newblock Automatic verification of pipelined microprocessor control.
\newblock In \emph{Computer Aided Verification: 6th International Conference,
  CAV'94 Stanford, California, USA, June 21--23, 1994 Proceedings 6}, 68--80.
  Springer.

\bibitem[{Burstall(1977)}]{burstall1977design}
Burstall, R.~M. 1977.
\newblock Design considerations for a functional programming language.
\newblock \emph{The software revolution}.

\bibitem[{Cadar et~al.(2008)Cadar, Ganesh, Pawlowski, Dill, and
  Engler}]{cadar2008exe}
Cadar, C.; Ganesh, V.; Pawlowski, P.~M.; Dill, D.~L.; and Engler, D.~R. 2008.
\newblock EXE: Automatically generating inputs of death.
\newblock \emph{ACM Transactions on Information and System Security (TISSEC)},
  12(2): 1--38.

\bibitem[{Christ, Hoenicke, and Nutz(2012)}]{smtinterpol}
Christ, J.; Hoenicke, J.; and Nutz, A. 2012.
\newblock SMTInterpol: An Interpolating SMT Solver.
\newblock In Donaldson, A.; and Parker, D., eds., \emph{Model Checking
  Software}, 248--254. Berlin, Heidelberg: Springer Berlin Heidelberg.
\newblock ISBN 978-3-642-31759-0.

\bibitem[{Clarke et~al.(2001)Clarke, Biere, Raimi, and Zhu}]{clarke2001bounded}
Clarke, E.; Biere, A.; Raimi, R.; and Zhu, Y. 2001.
\newblock Bounded model checking using satisfiability solving.
\newblock \emph{Formal methods in system design}, 19: 7--34.

\bibitem[{De~Angelis et~al.(2020)De~Angelis, Fioravanti, Pettorossi, and
  Proietti}]{DeAngelis2020chc}
De~Angelis, E.; Fioravanti, F.; Pettorossi, A.; and Proietti, M. 2020.
\newblock Removing Algebraic Data Types from Constrained Horn Clauses Using
  Difference Predicates.
\newblock In Peltier, N.; and Sofronie-Stokkermans, V., eds., \emph{Automated
  Reasoning}, 83--102. Cham: Springer International Publishing.
\newblock ISBN 978-3-030-51074-9.

\bibitem[{de~Moura and Bj{\o}rner(2008)}]{z3}
de~Moura, L.; and Bj{\o}rner, N. 2008.
\newblock Z3: An Efficient SMT Solver.
\newblock In Ramakrishnan, C.~R.; and Rehof, J., eds., \emph{Tools and
  Algorithms for the Construction and Analysis of Systems}, 337--340. Berlin,
  Heidelberg: Springer Berlin Heidelberg.
\newblock ISBN 978-3-540-78800-3.

\bibitem[{Ershov(1958)}]{hash-consing}
Ershov, A.~P. 1958.
\newblock On Programming of Arithmetic Operations.
\newblock \emph{Commun. ACM}, 1(8): 3–6.

\bibitem[{Goetz(2022)}]{javaADTs}
Goetz, B. 2022.
\newblock JEP 360: Sealed Classes (Preview).
\newblock \url{https://openjdk.org/jeps/360}.
\newblock Accessed: 2023-08-15.

\bibitem[{Gupta and Nau(1992)}]{gupta1992complexity}
Gupta, N.; and Nau, D.~S. 1992.
\newblock On the complexity of blocks-world planning.
\newblock \emph{Artificial intelligence}, 56(2-3): 223--254.

\bibitem[{Hoare(1975)}]{hoare1975recursive}
Hoare, C. A.~R. 1975.
\newblock Recursive data structures.
\newblock \emph{International Journal of Computer \& Information Sciences},
  4(2): 105--132.

\bibitem[{Hojjat and Rümmer(2017)}]{Hojjat}
Hojjat, H.; and Rümmer, P. 2017.
\newblock Deciding and Interpolating Algebraic Datatypes by Reduction.

\bibitem[{Hudak et~al.(2007)Hudak, Hughes, Peyton~Jones, and
  Wadler}]{hudak2007history}
Hudak, P.; Hughes, J.; Peyton~Jones, S.; and Wadler, P. 2007.
\newblock A history of Haskell: being lazy with class.
\newblock In \emph{Proceedings of the third ACM SIGPLAN conference on History
  of programming languages}, 12--1.

\bibitem[{Jung et~al.(2021)Jung, Jourdan, Krebbers, and Dreyer}]{jung2021safe}
Jung, R.; Jourdan, J.-H.; Krebbers, R.; and Dreyer, D. 2021.
\newblock Safe systems programming in Rust.
\newblock \emph{Communications of the ACM}, 64(4): 144--152.

\bibitem[{Kautz and Selman(1996)}]{kautz1996pushing}
Kautz, H.; and Selman, B. 1996.
\newblock Pushing the envelope: Planning, propositional logic, and stochastic
  search.
\newblock In \emph{Proceedings of the national conference on artificial
  intelligence}, 1194--1201.

\bibitem[{Kautz, Selman et~al.(1992)}]{kautz1992planning}
Kautz, H.~A.; Selman, B.; et~al. 1992.
\newblock Planning as Satisfiability.
\newblock In \emph{ECAI}, volume~92, 359--363. Citeseer.

\bibitem[{Kostyukov, Mordvinov, and Fedyukovich(2021)}]{Kostyukov}
Kostyukov, Y.; Mordvinov, D.; and Fedyukovich, G. 2021.
\newblock Beyond the elementary representations of program invariants over
  algebraic data types.

\bibitem[{Kov\'{a}cs, Robillard, and Voronkov(2017)}]{kovacs2017terms}
Kov\'{a}cs, L.; Robillard, S.; and Voronkov, A. 2017.
\newblock Coming to Terms with Quantified Reasoning.
\newblock In \emph{Proceedings of the 44th ACM SIGPLAN Symposium on Principles
  of Programming Languages}, POPL '17, 260–270. New York, NY, USA:
  Association for Computing Machinery.
\newblock ISBN 9781450346603.

\bibitem[{Lubarsky(2008)}]{lubarsky2008ian}
Lubarsky, R. 2008.
\newblock Ian Chiswell and Wilfrid Hodges. Mathematical logic. Oxford Texts in
  Logic, vol. 3. Oxford University Press, Oxford, England, 2007, 250 pp.
\newblock \emph{Bulletin of Symbolic Logic}, 14(2): 265--267.

\bibitem[{Milner(1997)}]{milner1997definition}
Milner, R. 1997.
\newblock \emph{The definition of standard ML: revised}.
\newblock MIT press.

\bibitem[{Oppen(1980)}]{Oppen1980}
Oppen, D.~C. 1980.
\newblock Reasoning About Recursively Defined Data Structures.
\newblock \emph{J. ACM}, 27(3): 403–411.

\bibitem[{Pham and Whalen(2014)}]{pham2014unrolling}
Pham, T.-H.; and Whalen, M.~W. 2014.
\newblock An Improved Unrolling-Based Decision Procedure for Algebraic Data
  Types.
\newblock In Cohen, E.; and Rybalchenko, A., eds., \emph{Verified Software:
  Theories, Tools, Experiments}, 129--148. Berlin, Heidelberg: Springer Berlin
  Heidelberg.
\newblock ISBN 978-3-642-54108-7.

\bibitem[{Reynolds and Blanchette(2017)}]{Reynolds2017codatatypes}
Reynolds, A.; and Blanchette, J.~C. 2017.
\newblock A Decision Procedure for (Co)datatypes in SMT Solvers.
\newblock \emph{Journal of Automated Reasoning}, 58(3): 341--362.

\bibitem[{Reynolds et~al.(2018)Reynolds, Viswanathan, Barbosa, Tinelli, and
  Barrett}]{Reynolds2018shared}
Reynolds, A.; Viswanathan, A.; Barbosa, H.; Tinelli, C.; and Barrett, C. 2018.
\newblock Datatypes with Shared Selectors.
\newblock In Galmiche, D.; Schulz, S.; and Sebastiani, R., eds.,
  \emph{Automated Reasoning}, 591--608. Cham: Springer International
  Publishing.
\newblock ISBN 978-3-319-94205-6.

\bibitem[{R{\"u}mmer and Wahl(2010)}]{rummer2010smt}
R{\"u}mmer, P.; and Wahl, T. 2010.
\newblock An SMT-LIB theory of binary floating-point arithmetic.
\newblock In \emph{International Workshop on Satisfiability Modulo Theories
  (SMT)}, 151.

\bibitem[{Rungta(2022)}]{rungta2022billion}
Rungta, N. 2022.
\newblock A billion SMT queries a day.
\newblock In \emph{International Conference on Computer Aided Verification},
  3--18. Springer.

\bibitem[{Russell(2010)}]{russell2010artificial}
Russell, S.~J. 2010.
\newblock \emph{Artificial intelligence a modern approach}.
\newblock Pearson Education, Inc.

\bibitem[{Salgado(2023)}]{pythonADT}
Salgado, P.~G. 2023.
\newblock What’s New In Python 3.10.
\newblock
  \url{https://docs.python.org/3.10/whatsnew/3.10.html#summary-release-highlights}.
\newblock Accessed: 2023-08-15.

\bibitem[{Sebastiani(2007)}]{Sebastiani2007lazy}
Sebastiani, R. 2007.
\newblock Lazy Satisfiability Modulo Theories.
\newblock 3: 141--224.

\bibitem[{Sussman(1973)}]{sussman1973computational}
Sussman, G.~J. 1973.
\newblock A computational model of skill acquisition.

\bibitem[{Suter, Dotta, and Kuncak(2010)}]{suter2010abstractions}
Suter, P.; Dotta, M.; and Kuncak, V. 2010.
\newblock Decision Procedures for Algebraic Data Types with Abstractions.
\newblock \emph{SIGPLAN Not.}, 45(1): 199–210.

\bibitem[{Winograd(1971)}]{winograd1971procedures}
Winograd, T. 1971.
\newblock Procedures as a representation for data in a computer program for
  understanding natural language.

\end{thebibliography}

\end{document}